\documentclass[twocolumn, twocolappendix]{aastex631}
\usepackage{amsmath}
\usepackage{enumitem}
\usepackage{newtxtext,newtxmath}
\usepackage{bm}

\hypersetup{breaklinks,colorlinks,citecolor=blue,urlcolor=blue,linkcolor=red}

\newcommand{\uatentry}[2]{\href{http://astrothesaurus.org/uat/#2}{#1 (#2)}}
\newcommand{\Msun}{\,{\rm M}_\odot}

\newcommand{\kms}{{\rm \,km\,s^{-1}}}

\newcommand{\E}{E}

\newcommand{\ecc}{e}

\graphicspath{{./}{figures/}}

\shorttitle{\it Particle spray algorithm for GC streams}
\shortauthors{\it Chen et al.}

\begin{document}

\title{\vspace{-6mm}\large Improved particle spray algorithm for modeling globular cluster streams}
\author[0000-0002-5970-2563]{Yingtian Chen}
\author[0000-0002-6257-2341]{Monica Valluri}
\author[0000-0001-9852-9954]{Oleg Y. Gnedin}
\author[0009-0003-7613-3109]{Neil Ash}
\affiliation{Department of Astronomy, University of Michigan, Ann Arbor, MI 48109, USA}

\correspondingauthor{Yingtian Chen}
\email{ybchen@umich.edu}

\begin{abstract}\noindent 
Stellar streams that emerge from globular clusters (GCs) are thin stellar structures spread along the orbits of progenitor clusters. Numerical modeling of these streams is essential for understanding their interaction with the host galaxy's mass distribution. Traditional methods are either computationally expensive or oversimplified, motivating us to develop a fast and accurate approach using a particle spray algorithm. By conducting a series of N-body simulations of GCs orbiting a host galaxy, we find that the position and velocity distributions of newly-escaped stream particles are consistent across various GC masses and orbital parameters. Based on these distributions, we develop a new algorithm that avoids computing the detailed internal cluster dynamics by directly drawing tracer particles from these distributions. This algorithm correctly reproduces the action space distribution of stream particles and achieves a 10\% accuracy in stream morphology and velocities compared to N-body simulations. To facilitate broader use, we have implemented this algorithm in galactic dynamics codes \texttt{agama}, \texttt{gala}, \texttt{galax}, and \texttt{galpy}.
\end{abstract}

\keywords{\uatentry{Stellar streams}{2166}; \uatentry{Globular star clusters}{656}; \uatentry{Galaxy dynamics}{591}; \uatentry{Galaxy structure}{622}; \uatentry{Computational astronomy}{293}; \uatentry{N-body simulations}{1083}}

\section{Introduction}

Stellar streams are the tidal debris of formerly bound progenitor stellar systems, such as globular clusters (GCs) and dwarf satellite galaxies \citep[e.g.,][]{lynden-bell_ghostly_1995}. Internal and external gravitational interactions cause the progenitor to eject stars beyond its Roche lobe via the $L_1$ and $L_2$ Lagrange points. In the progenitor's co-rotating frame, the Coriolis effect sprays these stars along the orbit, creating the trailing and leading arms of a stellar stream \citep[see e.g.,][]{binney_galactic_2008,weatherford_stellar_2024}.

Compared to dwarf galaxies, GCs have more compact mass profiles, with half-mass radii significantly smaller than the Roche lobe even at the orbital pericenter. Consequently, mass loss from a GC is governed by the gradual evaporation processes inside the cluster \citep[e.g.,][]{gnedin_destruction_1997,vesperini_effects_1997,baumgardt_dynamical_2003,weatherford_stellar_2023}, leading to velocity dispersion near the Lagrange points ($\lesssim1\kms$) that is much smaller than the orbital velocity ($\gtrsim100\kms$). Also, GCs are less sensitive to dynamical friction due to their lower masses. Therefore, GC streams such as GD-1 and Pal 5 are thinner structures that closely trace the orbits of their progenitors, providing strong constraints on the mass distribution of the host galaxy's dark matter halo \citep[e.g.,][]{koposov_constraining_2010,kupper_globular_2015}. Additionally, GC streams are more sensitive to gravitational perturbations from the galaxy's substructure since they are dynamically colder. For instance, a close encounter with a dark matter subhalo, another GC, or a giant molecular cloud can create density gaps in the stream \citep[e.g.,][]{carlberg_pal_2012,ngan_using_2014,erkal_number_2016,erkal_sharper_2017,banik_probing_2018}. However, such gaps may also be created by the rotation of galactic bars if the stream has a pericenter radius $\lesssim10$~kpc \citep[e.g.,][]{pearson_gaps_2017,bonaca_variations_2020}. To disentangle the multiple processes that alter the properties of GC streams, we must develop an \textit{accurate} theoretical method to understand how GC streams respond to their environment.

In addition to studying the properties of individual GC streams, joint analysis of multiple streams places even stronger constraints on the shape of the dark matter halo \citep[e.g.,][]{bovy_shape_2016}. Comprehensive measurements of star positions, kinematics, and photometry by the \textit{Gaia} mission \citep{gaia_collaboration_gaia_2016,gaia_collaboration_gaia_2018,gaia_collaboration_gaia_2023} over the past decade have revealed approximately $\sim100$ stellar streams in the Milky Way (MW), see a recent review by \citet{bonaca_stellar_2025}. Among these streams, around 80 are likely from GCs or fully disrupted GCs \citep{mateu_galstreams_2023,pearson_forecasting_2024}. By fitting the orbits of the thin and cold streams from \textit{Gaia}, \citet{ibata_charting_2024} constrained the virial mass of the MW to less than $20\%$ error. Nevertheless, future instruments like the \textit{Vera C. Rubin Observatory} and the \textit{Nancy Grace Roman Space Telescope} have the potential to reveal up to $\sim1000$ GC streams, which remain undiscovered today due to their low surface brightness and entangled morphology \citep{pearson_forecasting_2024}. Therefore, a \textit{fast} theoretical tool is necessary to model the entire GC stream population with affordable computational resources.

Unfortunately, developing a theoretical method that is both \textit{accurate} and \textit{fast} is challenging. To model the ejection of stars from the progenitor GC explicitly, we must simulate the gravitational interactions within the cluster using collisional N-body codes such as \texttt{NBODY6} \citep{aarseth_nbody1_1999} and \texttt{PeTar} \citep{wang_petar_2020}. While accurate, these codes are extremely computationally expensive due to the short time steps required to resolve close stellar encounters. Collisionless N-body codes avoid close encounters by employing a non-zero softening length. Although faster than collisional codes, they are still only suitable for a small number of stream realizations \citep[e.g.,][]{carlberg_simulating_2022,carlberg_subhalo_2023}.

To overcome the limitations of N-body simulations, a common practice is to directly eject stream tracer particles near the Lagrange points and draw their positions and velocities from an approximate distribution function \citep[e.g.,][]{varghese_stellar_2011,lane_tidal_2012,kupper_more_2012,bonaca_milky_2014,gibbons_skinny_2014,fardal_generation_2015,roberts_stellar_2024}. Each particle represents either a single star or a stellar population, depending on the specific model. These models then evolve the tracer particles solely under an analytical galactic potential and sometimes the progenitor's potential. The approximate distribution function should be carefully chosen to match the stream morphology and kinematics from N-body simulations. Since tracer particles are directly sprayed outside the Roche lobe without calculating the internal dynamics inside the progenitor GC, these methods are much faster than N-body simulations and are commonly referred to as \textit{particle spray} algorithms.

However, current particle spray algorithms make simplifications to reduce the number of model parameters. For example, the \citet{gibbons_skinny_2014} model sprays particles exactly at the Lagrange points, while the \citet{fardal_generation_2015} model assumes particles have no radial motion. Some of these simplifications are too unrealistic to accurately reproduce the stream morphology and kinematics from simulations. Additionally, most algorithms were calibrated using dwarf galaxy streams, as they were developed before the burst of discoveries of GC streams by \textit{Gaia}. It may be problematic to transplant the model parameters calibrated with more diffuse and massive dwarf galaxy streams to GC streams.

To improve the accuracy of particle spray methods, we must explicitly understand and parameterize the mechanism of stellar escape within GCs. To achieve this goal, we run a suite of N-body simulations of GCs orbiting a realistic galactic potential. By varying the cluster mass, orbital radius, eccentricity, and inclination, we find that the stream particles escape the progenitor cluster with similar initial position and velocity distributions, although the mass loss rate may diverge for different simulation settings. These distributions can be robustly approximated by a multivariate Gaussian distribution in a spherical coordinate system centered on the progenitor cluster. Subsequently, we develop a new particle spray algorithm that draws tracer particles from the Gaussian distribution calibrated from N-body simulations. This new method yields a remarkably close match to N-body simulations in both the action space and the 6D phase space (position $+$ velocity). Moreover, the new particle spray algorithm requires the same input as most existing algorithms, allowing straightforward implementation in galactic dynamics software such as \texttt{agama} \citep{vasiliev_agama_2019}, \texttt{gala} \citep{price-whelan_gala_2017,price-whelan_adrngala_2024}, \texttt{galax} \citep{starkman_galacticdynamicsgalax_2024}, and \texttt{galpy} \citep{bovy_galpy_2015}.

This paper is structured as follows. First, we describe a set of N-body simulations to study the 6D phase space distributions of stream particles in \S\ref{sec:simulations}. In \S\ref{sec:spray}, we develop a new particle spray algorithm using an analytical expression to approximate these distributions. Next, we examine the performance of the new algorithm in \S\ref{sec:performance} by comparing the stream kinematics and morphology with N-body simulations. Finally, we summarize our results in \S\ref{sec:summary}. We also test the effects of different initial profiles of the progenitor cluster in Appendix~\ref{sec:W_test} and perform convergence tests on numerical parameters in Appendix~\ref{sec:convergence}.

\section{N-body simulations}
\label{sec:simulations}

Before introducing the new particle spray algorithm, we quantitatively study the 6D phase space distributions of newly-escaped stream particles around the progenitor GC by running a suite of N-body simulations that cover a broad spectrum of GC masses, orbital radii, eccentricities, and inclinations.

\subsection{Simulation setup}
\label{sec:setups}

We run the N-body simulations using a fast-multipole gravity solver \texttt{falcON} \citep{dehnen_very_2000,dehnen_hierarchical_2002}. Each simulation features a GC orbiting in a MW-like potential. We initialize the GC with the King model \citep{king_structure_1966} given by the distribution function $f\propto\exp(\varepsilon/\sigma^2)-1$ if $\varepsilon>0$ and $f=0$ otherwise, where $\varepsilon\equiv-E+\Phi(0)$ is the relative energy to the central potential. 

The radius at which the stellar density of the King model drops to zero is the outer radius $r_{\rm outer}$. We set the outer radius as the tidal radius $r_{\rm tid}$ at the cluster's initial position. For a spherically symmetric potential, the tidal radius $r_{\rm tid}$ at a galactocentric radius $R$ is given by
\begin{equation}
    \left(\frac{r_{\rm tid}}{R}\right)^3 = \frac{M}{M_{\rm g}(R)}\left(3-\frac{d\ln M_{\rm g}(R)}{d\ln R}\right)^{-1}
    \label{eq:r_tid}
\end{equation}
where $M$ is the mass of the cluster and $M_{\rm g}(R)$ is the galaxy mass enclosed within $R$. This definition assumes that the potential is spherically symmetric, which is approximately valid outside the galactic disk.

Moreover, $\sigma$ in the King model is determined jointly by $r_{\rm outer}$ and the dimensionless central potential $W\equiv[\Phi(r_{\rm outer})-\Phi(0)]/\sigma^2$. Larger $W$ values correspond to more centrally concentrated profiles, with $W\rightarrow \infty$ reducing to the isothermal sphere. We set $W=8$ to match on the average the observed $r_{\rm tid}$--$r_{\rm core}$ relation for MW GCs from the \citet{hilker_galactic_2019} catalog\footnote{\url{https://people.smp.uq.edu.au/HolgerBaumgardt/globular/}}, where $r_{\rm core}$ is the core radius where stellar density drops to half the central density, see Fig.~\ref{fig:r_tid_r_core}. As we show in Appendix~\ref{sec:W_test}, a broad range of $W=4-14$ does not significantly alter the phase space distributions of escaped particles.

The GC is placed in the three-component galactic potential \texttt{MWPotential2014} as described by \citet{bovy_galpy_2015}. This potential is composed of a power-law bulge with an exponential cutoff, a Miyamoto--Nagai disk \citep{miyamoto_three-dimensional_1975}, and a Navarro--Frenk--White dark matter halo \citep{navarro_structure_1996}.

\begin{figure}
    \centering
    \includegraphics[width=0.9\linewidth]{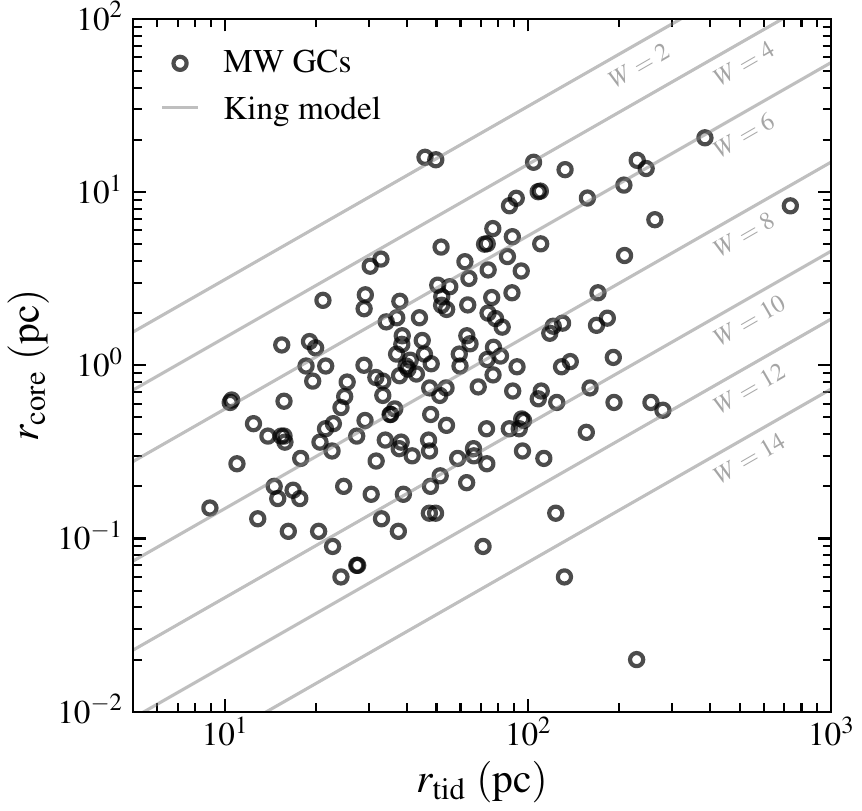}
    \caption{Tidal radius $r_{\rm tid}$ vs. core radius $r_{\rm core}$ for MW GCs from the \citet{hilker_galactic_2019} catalog. King models with the dimensionless central potential $W$ varying from 2 to 14 are shown as gray lines.}
    \label{fig:r_tid_r_core}
\end{figure}

We test three types of GC with initial masses $M_0=10^4\Msun$, $10^5\Msun$, and $10^6\Msun$, each represented by a collection of particles with mass $m=10\Msun$. We use the \texttt{falcON} default $P_1$ softening kernel with a softening length $\varepsilon=1$~pc. In Appendix~\ref{sec:convergence}, we verify that choosing different softening lengths ($0-4$~pc) and particle masses ($1.25-80\Msun$) does not change the phase space distributions of newly-escaped particles. We set the simulation time step $\tau=2^{-13}\ {\rm kpc\,km^{-1}\,s}\approx0.1$~Myr\footnote{The code unit of time is $\rm kpc\,km^{-1}\,s\approx0.978$~Gyr.}, which is chosen to be shorter than the typical softening length crossing time. Convergence tests in Appendix~\ref{sec:convergence} suggest that smaller time steps do not further improve integration accuracy.

\begin{deluxetable}{ccccc}
\tablecaption{Summary of orbital parameters employed in N-body simulations.}
\label{tab:nbody}
\tablehead{
\colhead{$R_{\rm apo}$} & \colhead{$R_{\rm peri}$} & \colhead{$\ecc$} & \colhead{$M_0$} & \colhead{Inclination} \\
$({\rm kpc})$ & $({\rm kpc})$ &  & $({\rm M_\odot})$ &  
}
\startdata
20 & 20 & 0 & $10^5$ & in-plane \\
20 & 10 & 0.3 & $10^5$ & in-plane \\
20 & 5 & 0.6 & $10^5$ & in-plane \\
\hline
40 & 40 & 0 & $10^4,10^5,10^6$ & in-plane \\
40 & 20 & 0.3 & $10^5$ & in-plane \\
40 & 10 & 0.6 & $10^4,10^5,10^6$ & in-plane \\
\hline
40 & 40 & 0 & $10^5$ & polar \\
40 & 10 & 0.6 & $10^5$ & polar \\
\enddata
\end{deluxetable} 

With $(\varepsilon,\tau)\approx(1\ {\rm pc},0.1\ {\rm Myr})$ we cannot resolve the close encounters inside the core of the cluster where the violent exchange of energy between stars is essential to eject stars outside the GC. In Appendix~\ref{sec:softening}, we use the collisional N-body code \texttt{PeTar} to simulate the close encounters without any softening. We find that the collisional simulation produces a slightly higher mass loss rate but almost identical phase space distributions of newly-escaped particles compared to our default settings. This is likely because the mass loss in the simulated GCs is dominated by the gradual evaporation processes rather than violent ejections \citep[see also][]{weatherford_stellar_2023}. Therefore, our \texttt{falcON} simulations are sufficient to correctly reveal the escape dynamics of stream particles.

We initialize the N-body cluster at the orbital apocenter in the disk plane, with $R_{\rm apo}\in\{20,40\}$~kpc. By varying the initial velocity, we obtain orbital eccentricities $\ecc\in\{0,0.3,0.6\}$, covering the typical eccentricity range of GC streams $\ecc=0.3-0.6$ \citep{li_s_2022}. These eccentricity values correspond to $R_{\rm apo}/R_{\rm peri}\in\{1,2,4\}$, yielding a galactocentric radius coverage of $5-40$~kpc, which includes most observed GC streams such as GD-1 and Pal~5.

\begin{figure*}
    \centering
    \includegraphics[width=\linewidth]{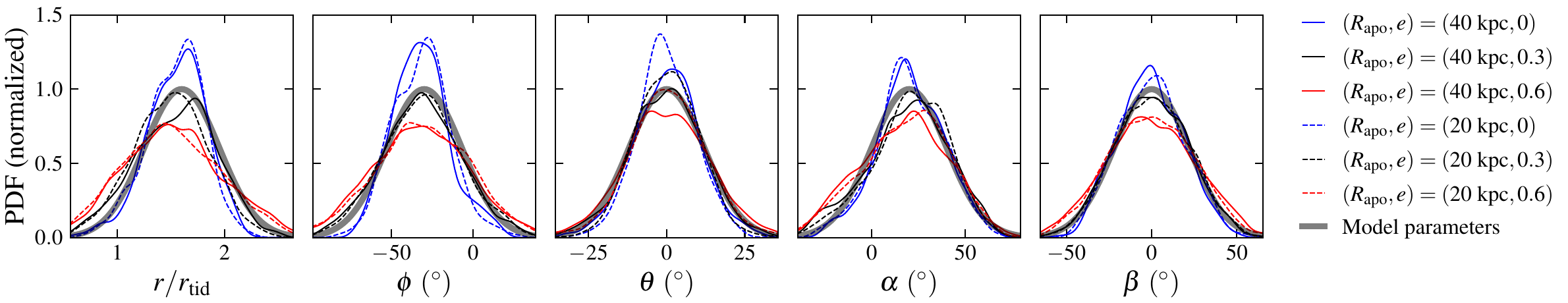}
    \caption{Initial position and velocity distributions for escaped particles from N-body simulations with fixed $M_0=10^5\Msun$ and varying $R_{\rm apo}$ and $\ecc$. We show our particle spray model distributions as thick gray curves. The probability distribution functions (PDFs) are obtained from the Gaussian KDE method, with bandwidths set to $1/5$ of our model standard deviations. We normalize the PDFs such that the peak values of our model distributions are unity.}
    \label{fig:spray_params_fixed_mean_1d}
\end{figure*}

\begin{figure*}
    \centering
    \includegraphics[width=\linewidth]{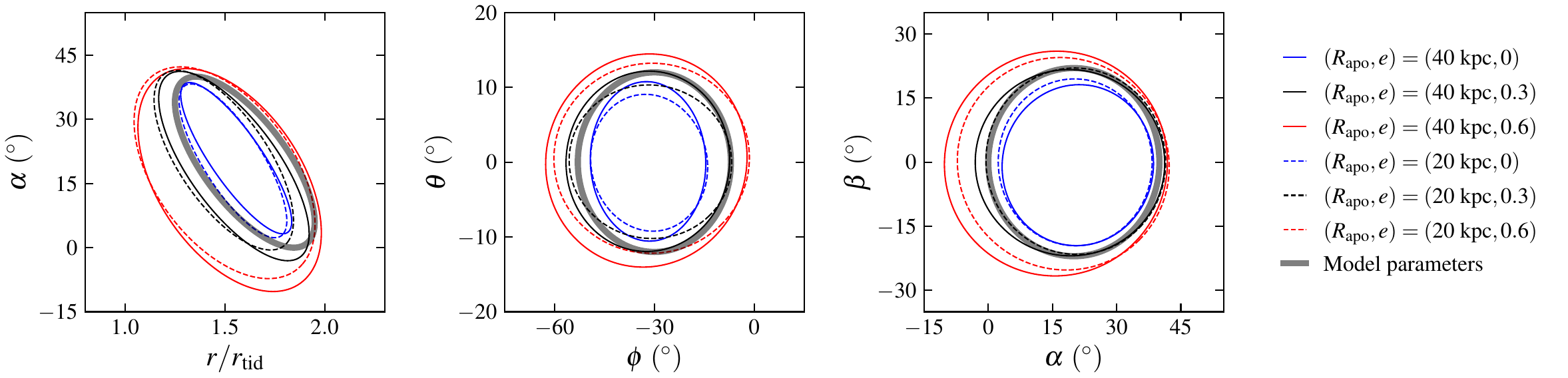}
    \caption{Initial position and velocity distributions for escaped particles as in Fig.~\ref{fig:spray_params_fixed_mean_1d}, but shown in 2D planes of 3 pairs of parameters. For clarity, we only plot the 1-$\sigma$ ellipses fit to all escaped particles in the $r$--$\alpha$ space (\textit{left}), $\phi$--$\theta$ space (\textit{middle}), and $\alpha$--$\beta$ space (\textit{right}). We also show the parameters of our particle spray model as thick gray ellipses.}
    \label{fig:spray_params_fixed_mean}
\end{figure*}

By default, all orbits have initial velocities within the disk plane. To study the impact of orbital inclination, we add two additional simulations with initial velocities rotated perpendicular to the disk plane for the $\ecc=0$ and $0.6$ cases. We refer to these additional simulations as the polar orbits to distinguish them from the default in-plane orbits.

In total, there are eight different cluster orbits across a wide range of cluster masses, orbital radii, eccentricities, and inclinations. Table~\ref{tab:nbody} lists the parameters used in the N-body simulations conducted in this work. We run each simulation for 3~Gyr. This simulation duration allows the cluster to complete multiple orbits around the galactic center, even for the longest orbital period of $\sim 1.4$~Gyr from $(R_{\rm apo}, \ecc) = (40\ {\rm kpc}, 0)$, ensuring that the entire orbit is fully sampled at least twice. At the end of each simulation, $5-50\%$ of the progenitor cluster's initial mass becomes gravitationally unbound. Additionally, we explore the case of streams from a fully disrupted globular cluster by manually removing the cluster in \S\ref{sec:fully_disrupted}.

\subsection{Phase space distribution of escaped particles}
\label{sec:distribution}

In this subsection, we quantitatively study the 6D position and velocity distributions of the newly-escaped particles.

The escaped particles from the N-body cluster are determined via an iterative process. In the first iteration, we compute the total energy of a particle as the sum of its gravitational energy induced by the other particles, and its kinetic energy relative to the center of mass of all particles. Particles with positive energy are treated as unbound and are excluded in subsequent calculations of the gravitational energy and the center of mass. We repeat the process for the remaining particles until the list of bound particles converges. A particle is defined as escaped when its energy increases to zero and remains non-negative until the end of the simulation.

An alternative definition of escape considers when a particle's kinetic energy reaches the effective potential at the Lagrange points. However, this energy threshold does not necessarily guarantee escape unless the particle moves directly toward the Lagrange points. We have verified that most particles continue to bounce within the cluster even when their energy exceeds this threshold, as pointed out by \citet{claydon_spherical_2019}. Furthermore, once a particle escapes through the Lagrange points, it quickly surpasses the zero-energy threshold in around the crossing time inside the cluster. In this case, the two definitions show little difference. Therefore, we consistently use the original zero-energy threshold definition throughout this work.

We use an inertial (i.e., non-rotating) frame constructed on the center of the cluster $\bm{R}_{\rm c}$\footnote{Throughout this work, we use upper-case coordinates in the galactocentric frame and lower-case coordinates in the cluster-centered frame.}, with the $x$--axis pointing away from the galactic center. The $z$--axis aligns with the orbital angular momentum vector of the cluster, such that the orbital velocity vector $\bm{V}_{\rm c}$ lies within the $x$--$y$ plane. The orientation of the $y$--axis is determined using the right-hand rule. We find it useful to describe the relative positions of escaped particles $\bm{r}\equiv\bm{R}\bm{-}\bm{R}_{\rm c}$ using a spherical coordinate system $(r,\phi,\theta)$, where $r$ is the distance of the particle from the cluster center, $\phi$ is the azimuthal angle from the $x$--axis, and $\theta$ is the latitude angle from the $x$--$y$ plane.

The relative velocity ${\bm v}\equiv{\bm V}\bm{-}\bm{V}_{\rm c}$ is described using the same coordinate system. For clarity, we use the symbols $(v,\alpha,\beta)$ to distinguish from the positional coordinates. Note that the inertial coordinate system is not co-rotating with the cluster, so ${\bm v}=\bm{\dot{r} + \Omega\times r}$. Here, $\bm{\Omega}$ is the orbital frequency of the cluster around the galaxy.

The phase space has six coordinates $(r,\phi,\theta,v,\alpha,\beta)$. These parameters are not fully independent. By definition, a newly-escaped particle has near-zero total energy, so its velocity approximately equals the escape velocity at the current position. Since most particles escape around the tidal radius, which is typically much larger than the half-mass radius of the cluster, their velocity can be well approximated by the escape velocity of a point-mass potential $v_{\rm esc}\equiv\sqrt{2GM/r}$, with a deviation $<4\%$. Therefore, we can effectively treat $v$ as a one-to-one function of $r$, leaving only five independent coordinates $(r,\phi,\theta,\alpha,\beta)$. 

Figure~\ref{fig:spray_params_fixed_mean_1d} shows the distributions of the five position and velocity parameters for in-plane orbits with varying $R_{\rm apo}$ and $\ecc$. We obtain the distributions by Gaussian kernel density estimation (KDE). Here, we fix the cluster mass to $10^5\Msun$. All angles are given in degrees, and $r$ is normalized by the tidal radius $r_{\rm tid}$ at the current galactocentric radius of the GC. The actual distributions of azimuthal angles $\phi$ and $\alpha$ are bimodal, corresponding to the trailing and leading stream arms. For visual clarity, we shift the azimuth angles of the leading arm ($\phi>60^\circ$ or $\phi<-120^\circ$) by $180^\circ$ such that the two arms are located in the same region. We notice that the distributions of trailing and leading arms are indistinguishable after the shift, indicating that the initial escape positions and velocities are symmetric in both the $\phi$ and $\alpha$ coordinates.

We find that these distributions approximately follow Gaussian distributions. While $\theta$ and $\beta$ are symmetric around zero, the other parameters are centered around $\overline{r} = (1.5-1.6)\,r_{\rm tid}$, $\overline{\phi} \approx -30^\circ$, and $\overline{\alpha} \approx 20^\circ$. The mean escape radius lies outside the tidal radius because our escape energy threshold is slightly higher than the minimum energy required for escape. This ensures that escaping particles interact minimally with the progenitor and leave the cluster immediately. This feature is important for developing a stable particle spray model.

\begin{figure}
    \centering
    \includegraphics[width=\linewidth]{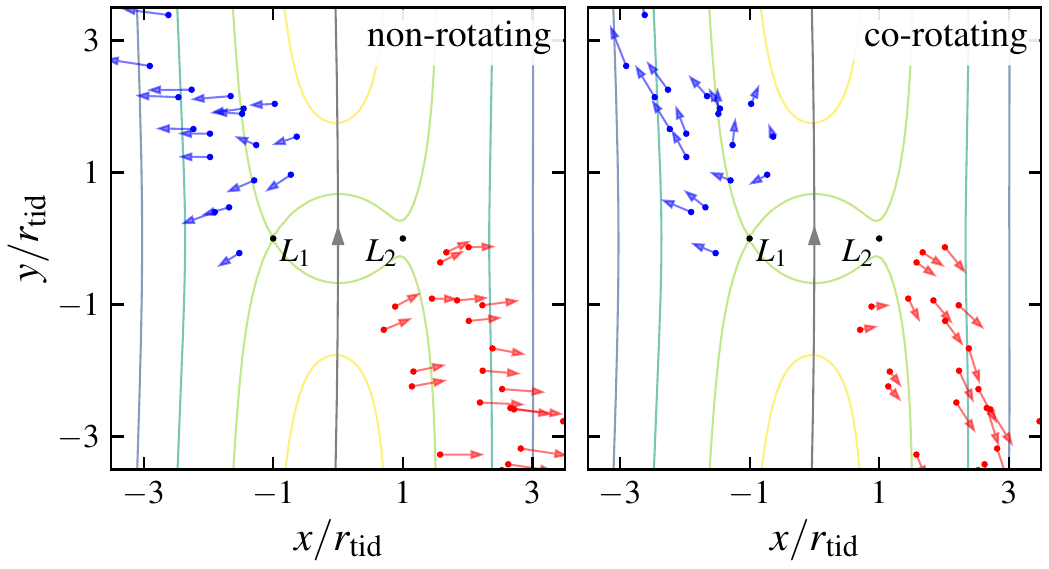}
    \caption{Positions and velocity vectors of escaped particles near the $L_1$ and $L_2$ points at the last simulation output with $(M_0, R_{\rm apo}, \ecc)=(10^5\Msun,20\ {\rm kpc}, 0)$. We show the velocity vectors in the non-rotating frame (\textit{left}) and co-rotating frame (\textit{right}) centered on the progenitor cluster. Trailing (red) and leading (blue) particles are plotted in different colors, with background contours showing the effective potential around the cluster. An animated version of this figure is available in the online journal and can also be accessed at \url{https://github.com/ybillchen/particle_spray}. The animation shows particles escaping over 1~Gyr before the presented snapshot.}
    \label{fig:escape_velocity}
\end{figure}

To study the correlations between the five parameters, we plot the 1-$\sigma$ ellipses of these parameters in Fig.~\ref{fig:spray_params_fixed_mean}. Most parameters are uncorrelated, except that $\alpha$ anti-correlates with $r$, where $\alpha$ drops from $\sim60^\circ$ to $0^\circ$ when $r$ increases from $r_{\rm tid}$ to $2r_{\rm tid}$. To demonstrate this behavior, we plot in the \textit{left panel} of Fig.~\ref{fig:escape_velocity} the positions and velocity vectors of escaped particles near $L_1$ and $L_2$ for one of the simulation runs at the last output. We find that particles closer to the cluster center tend to escape with a stronger prograde motion. The positive $\alpha$ is in agreement with the findings of most existing particle spray algorithms \citep[e.g.,][]{fardal_generation_2015, roberts_stellar_2024}. Subsequently, the Coriolis force quickly shifts $\alpha$ to negative values in the co-rotating frame (with angular frequency $\bm{\Omega}$) centered on the progenitor cluster, as shown in the \textit{right panel} of Fig.~\ref{fig:escape_velocity}.

The mean values of these Gaussian distributions are almost invariant with respect to $R_{\rm apo}$ or $\ecc$, although orbits with higher eccentricities tend to have larger variations around the means. As $\ecc$ increases from 0 to 0.6, the variations of all coordinate parameters also increase by a factor of $\sim2$. This is likely due to the stronger tidal heating in more eccentric orbits, leading to more divergent escape positions and velocities. In contrast, $R_{\rm apo}$ weakly influences the variations of all coordinates.

\begin{figure*}
    \centering
    \includegraphics[width=\linewidth]{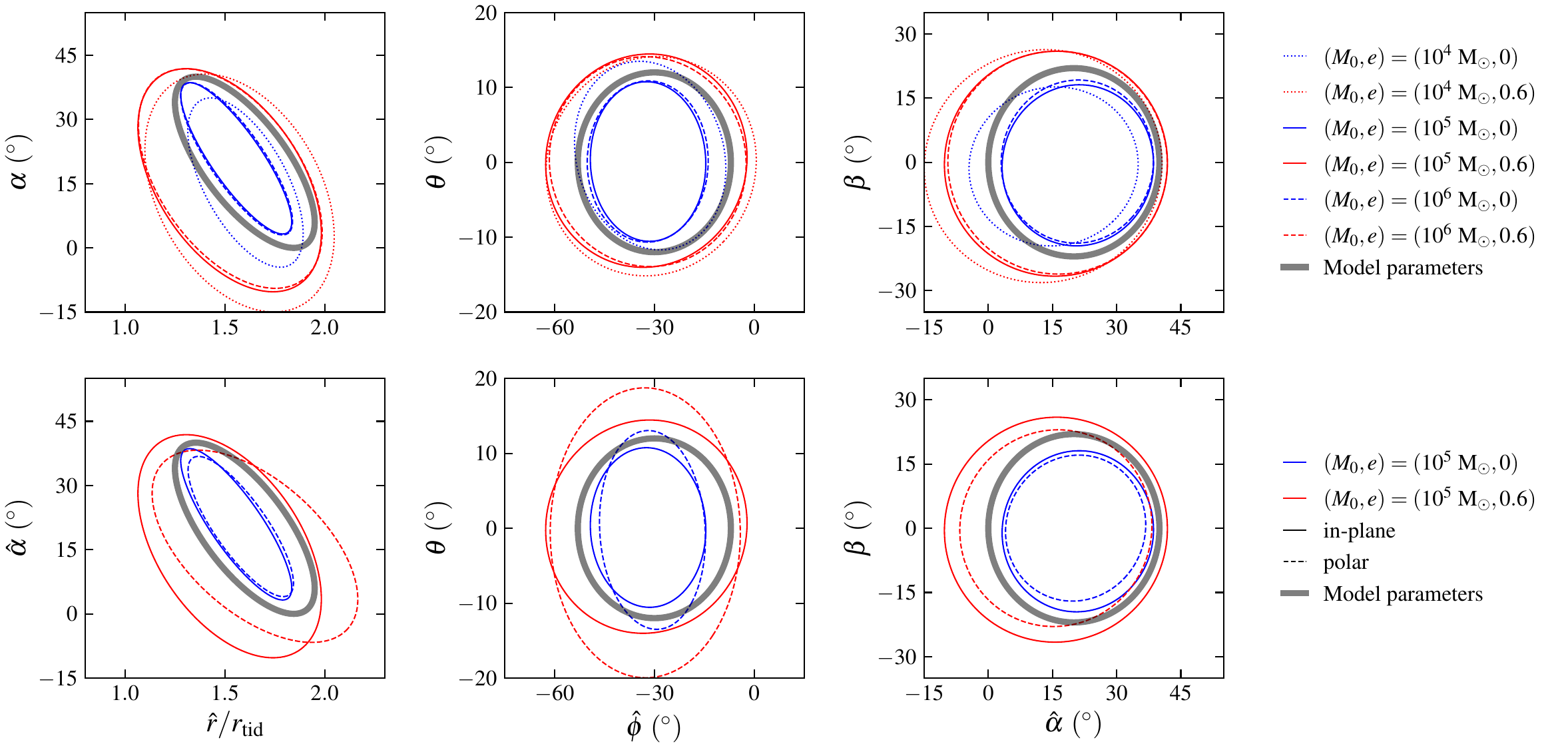}
    \caption{Initial position and velocity distributions for escaped particles as in Fig.~\ref{fig:spray_params_fixed_mean}, but with fixed $R_{\rm apo}=40$~kpc and varying $M_0$ and $\ecc$ (\textit{top row}). We also show the distributions for polar and in-plane orbits in the \textit{bottom row}.}
    \label{fig:spray_params_cluster_mass_orientation}
\end{figure*}

In Fig.~\ref{fig:spray_params_cluster_mass_orientation}, we show the phase space distributions for different $M_0$ and orbital inclinations. Since eccentricity may influence the distributions, we also vary $\ecc\in\{0,0.6\}$ to study the different behaviors for the circular and eccentric orbits. For clarity, we fix $R_{\rm apo}=40$~kpc as $R_{\rm apo}$ only weakly alters the distributions. We notice that the initial cluster mass $M_0$ does not significantly change the phase space distributions for $M_0\geq10^5\Msun$. However, the $(M_0,\ecc)=(10^4\Msun,0)$ case tends to have more scattered phase space distributions, likely due to the N-body noise when the number of particles is small, with only about 100 particles eventually stripped by the end of the simulation.

Moreover, the polar and in-plane cases yield almost identical parameter distributions for the circular orbits $\ecc=0$. For the eccentric orbits $\ecc=0.6$, however, the polar case tends to increase the mean radius to $\overline{r} \approx 1.7\,r_{\rm tid}$ and produce larger variation in $\theta$. This is likely due to the tidal shock when the cluster passes the disk plane near the pericenter, which is only 10~kpc from the galactic center. In contrast, the polar cases yield slightly smaller variation in $\phi$, $\alpha$, and $\beta$

\section{New particle spray algorithm}
\label{sec:spray}

A typical particle spray algorithm involves three steps. The first step is to integrate (either forward or backward) the progenitor's orbit $\bm{R}_{\rm c}(t)$ in an analytical galactic potential. Next, for each differential time $dt$, we generate the initial positions and velocities of $\dot{N}(t)dt$ tracer particles following the joint distribution function $f(\bm{r},\bm{v})$. Finally, we integrate the orbits of these particles in the combined gravitational potential of the host galaxy and the progenitor GC.

The particle spray algorithm takes the following inputs: the functional form of the galactic potential; the initial position and velocity of the progenitor cluster; and the mass history $M(t)$ of the cluster. The last input is needed when all tracer particles have the same mass $m$, such that $\dot{N}(t)=\dot{M}(t)/m$. For most existing particle spray algorithms, the tidal radius $r_{\rm tid}(R_{\rm c},M)$ is also required in the distribution function. In this case, $M(t)$ is also used to compute $r_{\rm tid}$. 

The defining characteristic of a particle spray model is the distribution function $f$. In most existing models, the distribution function either has a fixed functional form of $f(\bm{r},\bm{v})$ at different orbital phases or depends solely on $r_{\rm tid}$. However, knowing only $r_{\rm tid}$ is insufficient to fully capture the different orbital properties of the progenitor.

Fortunately, we find that the distributions of the five independent coordinates $(r,\phi,\theta,\alpha,\beta)$ for newly-escaped particles depend weakly on the orbital radius, inclination, and eccentricity (see \S\ref{sec:simulations}). If we normalize the escape radius $r$ by $r_{\rm tid}$, the mass of the cluster also does not influence the distributions. On the other hand, the variations of these coordinates tend to increase by a factor of $\sim2$ when changing the orbital eccentricity from $\ecc=0$ to $0.6$. The orbital inclination also slightly shifts the variations. However, these are only minor changes that do not significantly affect the general shapes of the distributions.

The robustness of these distributions allows for a simple and universal functional form of the distribution function that works for various types of orbits. Inspired by the Gaussian-like distributions of the five coordinates, we introduce a new particle spray algorithm using a multivariate Gaussian distribution function:
\begin{equation}
    f(r, \phi, \theta, \alpha, \beta)
    = {\cal N}(\bm{\mu}, \bm{\Sigma}) \quad \text{and} \quad v = v_{\rm esc}(r)
    \label{eq:gaussian}
\end{equation}
where $\bm{\mu}\equiv(\overline{r}, \overline{\phi}, \overline{\theta}, \overline{\alpha}, \overline{\beta})$ is the mean vector, and
\begin{equation}
    \bm{\Sigma}\equiv\left(
    \begin{array}{ccccc}
        \sigma_r^2 & & & C_{r\alpha} & \\
         & \sigma_\phi^2 & & & \\
         & & \sigma_\theta^2 & & \\
        C_{r\alpha} & & & \sigma_\alpha^2 & \\
         & & & & \sigma_\beta^2 \\
    \end{array}
    \right)
\end{equation}
is the covariance matrix. $C_{r\alpha}\equiv R_{r\alpha}\sigma_r\sigma_\alpha$ is the covariance between $r$ and $\alpha$, where $R_{r\alpha}$ is the Pearson correlation coefficient. We find that the following parameters approximately describe the phase space distributions for the various orbital types considered in this work:
\begin{equation}
\begin{aligned}
    & (\overline{r}, \overline{\phi}, \overline{\theta}, \overline{\alpha}, \overline{\beta}) \\
    &\quad=\ (1.6 r_{\rm tid}, -30^\circ, 0^\circ, 20^\circ, 0^\circ); \\
    & (\sigma_r, \sigma_\phi, \sigma_\theta, \sigma_\alpha, \sigma_\beta, R_{r\alpha}) \\
    &\quad=\ (0.35 r_{\rm tid}, 23^\circ, 12^\circ, 20^\circ, 22^\circ, -0.7).
    \label{eq:parameter}
\end{aligned}
\end{equation}
For comparison, we plot the 1-$\sigma$ ellipses from these parameters in Figs.~\ref{fig:spray_params_fixed_mean} and \ref{fig:spray_params_cluster_mass_orientation}. Although not perfect matches for all orbital types, these ellipses capture the main features of the position and velocity parameter distributions across a wide range of $M_0$, $R_{\rm apo}$, $\ecc$, and orbital inclinations. For example, we select $\overline{r}=1.6 r_{\rm tid}$ between the observed mean values from the in-plane ($\overline{r}=(1.5-1.6)r_{\rm tid}$) and polar ($\overline{r}\approx1.7r_{\rm tid}$) orbits. However, as we show in \S\ref{sec:performance}, our settings can still provide a good match to streams from N-body simulations even for the $\ecc=0.6$ cases, which seem to have much larger variation than our model parameters in Eq.~(\ref{eq:parameter}).

\section{Algorithm performance}
\label{sec:performance}

In this section, we assess the accuracy of the particle spray algorithm presented in \S\ref{sec:spray} by comparing the morphology and kinematics of individual streams created by this algorithm with those from N-body simulations. To examine whether our new method can be applied to a broader range of galactic potential models beyond \texttt{MWPotential2014}, we run a new suite of N-body simulations under the \citet{mcmillan_mass_2017} potential, whose circular velocity is greater than that of \texttt{MWPotential2014} by up to 15\% at $R=5-40$~kpc.

We test different orbital types of the progenitor cluster with $\ecc\in\{0,0.6\}$ in either in-plane or polar orbits. Similarly to \S\ref{sec:setups}, we initialize the cluster at the apocenter and vary its velocity vector to obtain different eccentricities and inclinations. In this section, we only present the runs with $R_{\rm apo}=40$~kpc and $M_0=10^5\Msun$, while our conclusions for other $R_{\rm apo}$ and $M_0$ values are similar.

In addition, we compare our algorithm with the following two commonly-used particle spray algorithms.
\begin{itemize}[labelindent=0pt,labelwidth=10pt,labelsep*=0pt,leftmargin=!,align=parleft]
    \item The \citet{fardal_generation_2015} model constructs a cylindrical galactocentric coordinate system $(R, Z, \phi)$, where $R$ represents the distance from the galactic center, and the $Z$--axis is aligned with the progenitor cluster's orbital angular momentum. Therefore, the $Z$--axis is perpendicular to the cluster's orbital plane. $\phi$ is the azimuth angle within the orbital plane. Escaped particles are ejected at $R = R_{\rm c} \pm k_r r_{\rm tid}$ with $\phi = \phi_{\rm c}$. The $Z$--coordinate follows a univariate Gaussian distribution $Z \sim \mathcal{N}(0, 0.5\,r_{\rm tid})$. The escape velocity is purely tangential, i.e., $V_R=0$. The azimuthal velocity $V_\phi=V_{\phi,{\rm c}}\pm k_{vt} V_{\rm circ}r_{\rm tid}/R_{\rm c}$; and the $Z$-component velocity $V_Z\sim{\cal N}(0,0.5\,V_{\rm circ}r_{\rm tid}/R_{\rm c})$. The values of $k_r$ and $k_{vt}$ are also drawn from Gaussian distributions with $\overline{k}_r=2$ and $\overline{k}_{vt}=0.3$. The standard deviations of the two random variables are both $\sigma=\min(0.15f_{\rm t}^2R_{\rm acc}^{2/3},0.4)$, where $f_{\rm t}$ is the ratio between the outer radius of the King model and $r_{\rm tid}$. Here, we have $f_{\rm t}=1$ by definition, see \S\ref{sec:setups}. $R_{\rm acc}$ is the ratio between the ``acceleration gradients'' \citep[defined in Eq.~(14) of][]{fardal_generation_2015} at orbital pericenter and apocenter. The above means and deviations are optimized by eye to match N-body simulations.

    A variation of the \citet{fardal_generation_2015} model sets $\sigma=0.5$ (e.g., \texttt{gala} version$<$1.9). This variation has been extensively used and has also led to promising results. We also test this setting but find no significant difference from the original model. Therefore, we do not include this variation in our comparison.

    \item The QSG-PS model by \citet{roberts_stellar_2024} constructs a Cartesian coordinate system centered on the galactic center with the $X$--axis pointing towards the star cluster. The $Z$--axis aligns with the cluster's orbital angular momentum such that the velocity vector is in the $X$--$Y$ plane. This model directly ejects escaped particles with a $\pm f_{\rm e}r_{\rm tid}$ offset from the cluster center along the $X$--axis. The initial velocities are isotropic with dispersion given by Eq.~(10) of \citet{roberts_stellar_2024} in all directions. In addition, the $Y$-direction velocity has a co-rotating component around the cluster quantified by $0\leq\epsilon\leq 1$, where $\epsilon=0$ stands for the escaped particles having on average the cluster's orbital velocity, while $\epsilon=1$ means that the particles have on average the cluster's angular velocity. \citet{roberts_stellar_2024} have optimized the model parameters to $(f_{\rm e},\epsilon)=(1.5,0.57)$ for the best match to N-body simulations.
\end{itemize}

Both the \citet{fardal_generation_2015} and \citet{roberts_stellar_2024} models neglect the gravitational interaction between escaped particles and the progenitor star cluster, i.e., particles are only influenced by the host galaxy's potential.

There are several other popular models not mentioned here. However, these models require additional free parameters that are not predefined. For example, the \citet{gibbons_skinny_2014} model is very similar to the $(f_{\rm e},\epsilon)=(1,0)$ case of \citet{roberts_stellar_2024}, while the former sets the progenitor's velocity dispersion and scale length as free parameters. We therefore exclude it from this comparison as its performance depends on the specific choice of parameters.

The N-body simulations output positions $\bm{R}_{\rm c}(t)$, velocities $\bm{V}_{\rm c}(t)$, and mass $M(t)$ of the progenitor cluster in a sequence of snapshots. These values are then passed to the particle spray models as input (see \S\ref{sec:spray}). Although the \citet{fardal_generation_2015} model has its own prescription to determine the particle escape rate, we still enforce the same escape rate history $M(t)$ given by the corresponding N-body simulation so that we can systematically study the model differences only due to the particle spray prescriptions. Specifically, we linearly interpolate the cluster's mass history between snapshots from N-body simulations and eject a particle every time the cluster mass decreases by a particle mass $m$. The particle is set to either trail or lead the progenitor cluster with equal probability.

The default model parameters are given in \S\ref{sec:spray}. We always refer to this model as ``our model'' or ``default model'' throughout the paper unless specifically mentioned. Our default model accounts for the gravitational interaction with the progenitor cluster by computing the orbits of escaped particles in the joint potential of the galaxy and the cluster. We approximate the potential of the star cluster as a point-mass particle. However, we also examine the Plummer model $\Phi(r)=GM/\sqrt{r^2+a^2}$ with $a=0-4$~pc (where $a=0$ reduces to the point mass), and these choices do not significantly alter the kinematics and morphology of the generated streams since $a\ll r_{\rm tid}$.

To understand the importance of including the progenitor's potential, we include a model variation that only accounts for the potential of the host galaxy. We refer to this variation as the ``no progenitor'' model.

\subsection{Action space distribution}

\begin{figure*}
    \centering
    \includegraphics[width=0.7\linewidth]{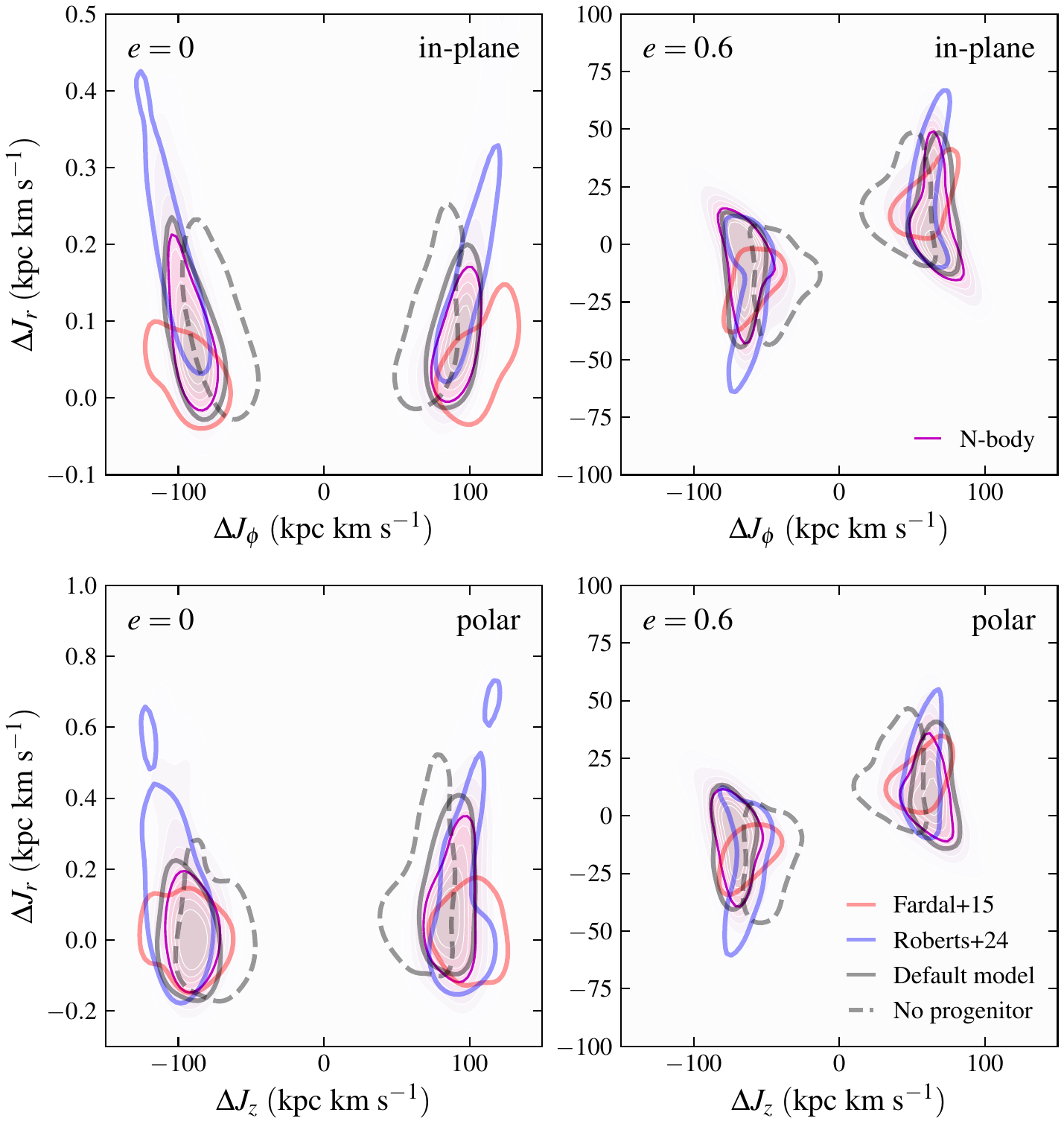}
    \caption{Distributions of stream particles in the action space relative to the progenitor cluster from circular (\textit{left column}) and eccentric (\textit{right column}) orbits with in-plane (\textit{top row}) and polar (\textit{bottom row}) inclinations. We fix $(M_0,R_{\rm apo})=(10^5\Msun,40\ {\rm kpc})$. For each panel, we show the distribution from N-body simulations as background contours, highlighting the contour enclosing 50\% of total particles as the thin magenta curve. We plot the 50\%-contours for particle spray models as thick curves.}
    \label{fig:action_comparison}
\end{figure*}

\begin{figure*}
    \centering
    \includegraphics[width=\linewidth]{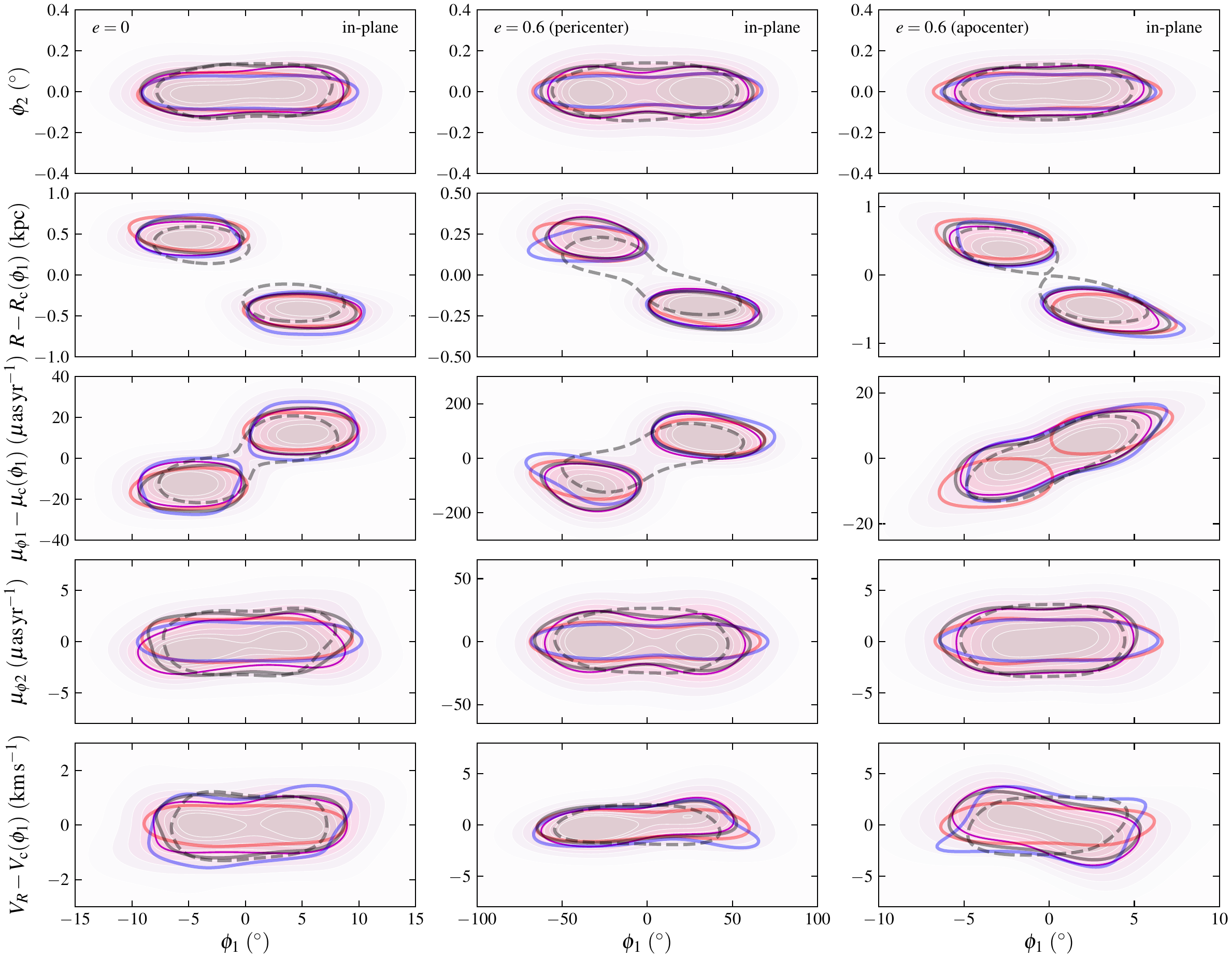}
    \caption{From \textit{top} to \textit{bottom}: stream morphology in the $\phi_1$--$\phi_2$ space (\textit{first row}) and $\phi_1$--$R$ space (\textit{second row}); and proper motions $\mu_{\phi 1}$ (\textit{third row}), $\mu_{\phi 2}$ (\textit{fourth row}), and radial velocity $V_R$ (\textit{fifth row}) plotted along $\phi_1$. We subtract the values of the progenitor cluster's orbits from $R$, $\mu_{\phi 1}$, and $V_R$ such that the cluster is always represented by zero. We fix $(M_0,R_{\rm apo})=(10^5\Msun,40\ {\rm kpc})$ and study in-plane orbits with $\ecc=0$ (\textit{left column}) and $\ecc=0.6$ (pericenter in the \textit{middle column} and apocenter in the \textit{right column}). The contours in each panel with different colors and linestyles refer to the same simulations/particle spray models as in Fig.~\ref{fig:action_comparison}.}
    \label{fig:morphology_compare}
\end{figure*}

\begin{figure*}
    \centering
    \includegraphics[width=\linewidth]{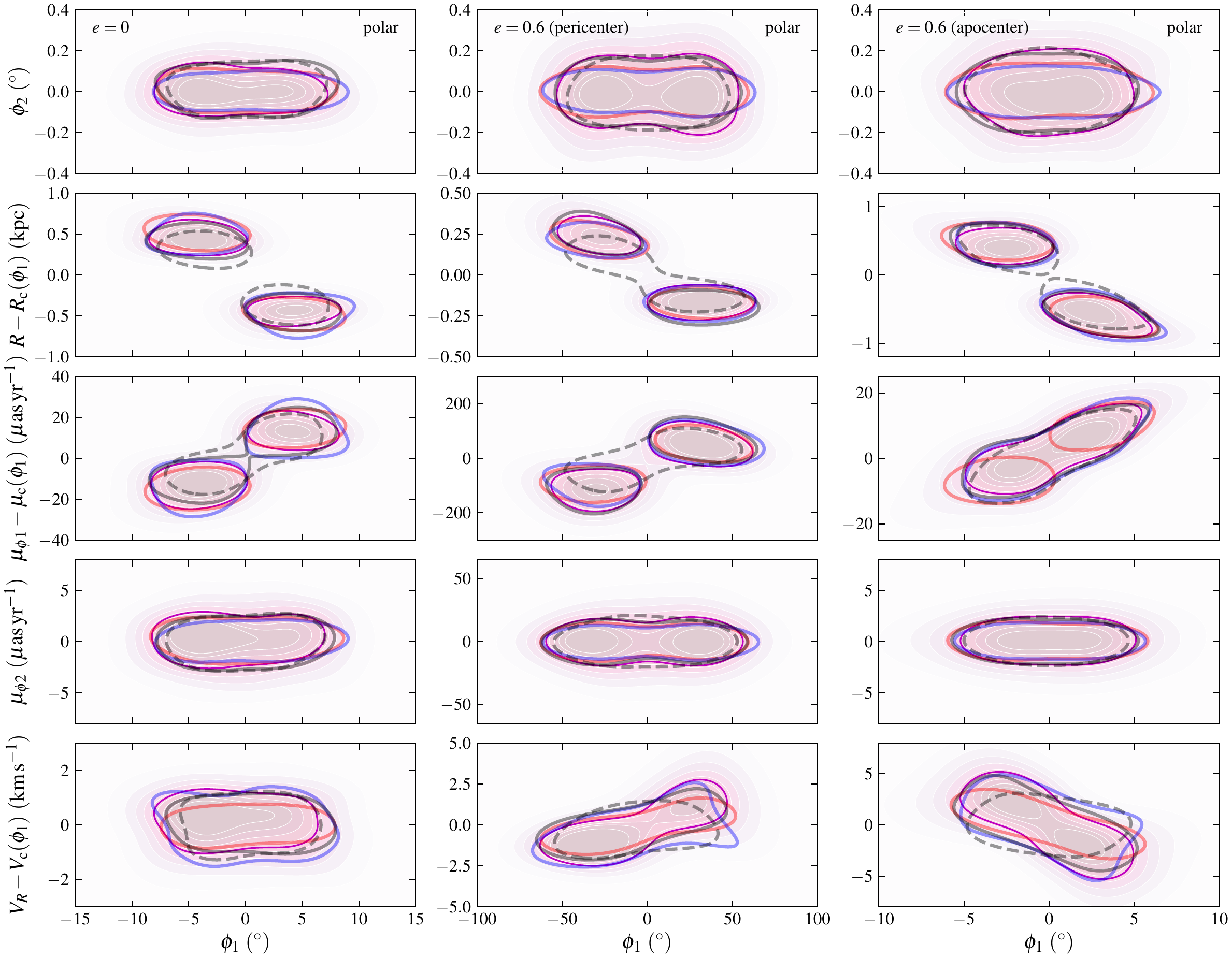}
    \caption{Stream morphology and velocities as in Fig.~\ref{fig:morphology_compare}, but for polar orbits.}
    \label{fig:morphology_compare_polar}
\end{figure*}

A common benchmark for particle spray recipes is to reproduce the orbital action space distribution of stream particles in N-body simulations. The orbital actions of a closed orbit are defined as
\begin{equation}
    J_q \equiv \frac{1}{2\pi} \oint v_q \, dq
    \label{eq:actions}
\end{equation}
where $q\in\{r,\phi,z\}$ are the radial, azimuthal, and vertical coordinates. We compute the actions using the code \texttt{agama} \citep{vasiliev_agama_2019}, which implements the St\"{a}ckel fudge method \citep{binney_actions_2012}. In an axisymmetric system, $J_\phi$ equals the angular momentum perpendicular to the galactic plane, $L_z$.

In Fig.~\ref{fig:action_comparison}, we compare the action space distribution of escaped particles relative to the actions of the progenitor cluster for our particle spray model with the N-body simulations and other particle spray models. For in-plane orbits, we have $J_z\approx0$ and thus only study the distributions in the $J_\phi$--$J_r$ space. In contrast, we have $J_\phi\approx0$ for polar orbits, where we focus on the $J_z$--$J_r$ space. We show the distribution at the last snapshot. The distributions at other snapshots are similar to the last snapshot since the actions are approximately conserved quantities as the particles primarily interact with the static, axisymmetric galactic potential. The distributions are represented by contours derived from Gaussian KDE.

We use the KL divergence \citep{kullback_information_1951} to quantify the similarity between distributions in the action space. The KL divergence between two distributions $p$ and $q$ is given by
\begin{equation*}
    D_{\rm KL}(p\|q)\equiv \int_{-\infty}^{\infty} p(x)\ln\frac{p(x)}{q(x)}dx.
    \label{eq:KL}
\end{equation*}
The greater the KL divergence, the easier it is to separate the two distributions. If $p$ and $q$ are identical, we have $D_{\rm KL}=0$. For reference, two 1D Gaussian distributions separated by 2 standard deviations have $D_{\rm KL}=2$. For our purpose, we set $p$ as the distribution from particle spray algorithms and $q$ as the corresponding N-body simulation. Numerically, we approximate the KL divergence using importance sampling:
\begin{equation*}
    D_{\rm KL}(p\|q)\approx \frac{1}{n}\sum_{x\in X} \frac{p(x)}{r(x)}\ln\frac{p(x)}{q(x)}
\end{equation*}
where $X$ is a set of $n$ points sampled from a distribution $r$. If we take $X$ to be the escaped particles from particle spray algorithms, which by construction are sampled from $p$, the expression simplifies to
\begin{equation*}
    D_{\rm KL}(p\|q)\approx \frac{1}{n}\sum_{x\in X} \ln\frac{p(x)}{q(x)},
\end{equation*}
where $p$ and $q$ are approximated by KDE.

For the circular orbits $\ecc=0$, the trailing and leading arms form two distinct lobes roughly symmetric about the line of $\Delta J_\phi=0$. Although producing slightly larger dispersion in $J_r$, our default model is still the best match to N-body simulations with $D_{\rm KL}<0.15$. In contrast, the \citet{fardal_generation_2015} model underestimates the dispersion of both lobes by a factor of up to 2, and the \citet{roberts_stellar_2024} model overestimates it by a factor of $\sim2$. These two models yield $D_{\rm KL}=1.6-2$ for the in-plane orbit and $D_{\rm KL}=0.6-1$ for the polar case.

For the eccentric orbit $\ecc=0.6$, the trailing and leading lobes are symmetric around the center point. Our default model is again the best match with $D_{\rm KL}<0.1$, whereas the other two models either systematically underestimate or overestimate the dispersion of the actions, producing a much greater $D_{\rm KL}=0.4-1$ for both the in-plane and polar orbits.

We emphasize that our model closely matches the action space distributions from N-body simulations, even though the initial positions and velocities for newly escaped particles differ notably for with different $\ecc$ (see Fig.~\ref{fig:spray_params_fixed_mean}) and orbital inclinations (see the \textit{bottom row} of Fig.~\ref{fig:spray_params_cluster_mass_orientation}). This indicates that one universal set of spray parameters is sufficient to reproduce the action space distributions of streams with a wide range of orbital properties.

We also note that the ``no progenitor'' version of our model yields trailing and leading lobes significantly closer in the $J_\phi$ or $J_z$ direction. The corresponding KL divergence increases to around the same level as those of \citet{fardal_generation_2015} and \citet{roberts_stellar_2024}, highlighting the importance of including the progenitor's potential in reproducing the action space distribution from N-body simulations. This has also been emphasized by \citet{gibbons_skinny_2014}.

\subsection{Morphology and velocities}

\begin{figure*}
    \centering
    \includegraphics[width=\linewidth]{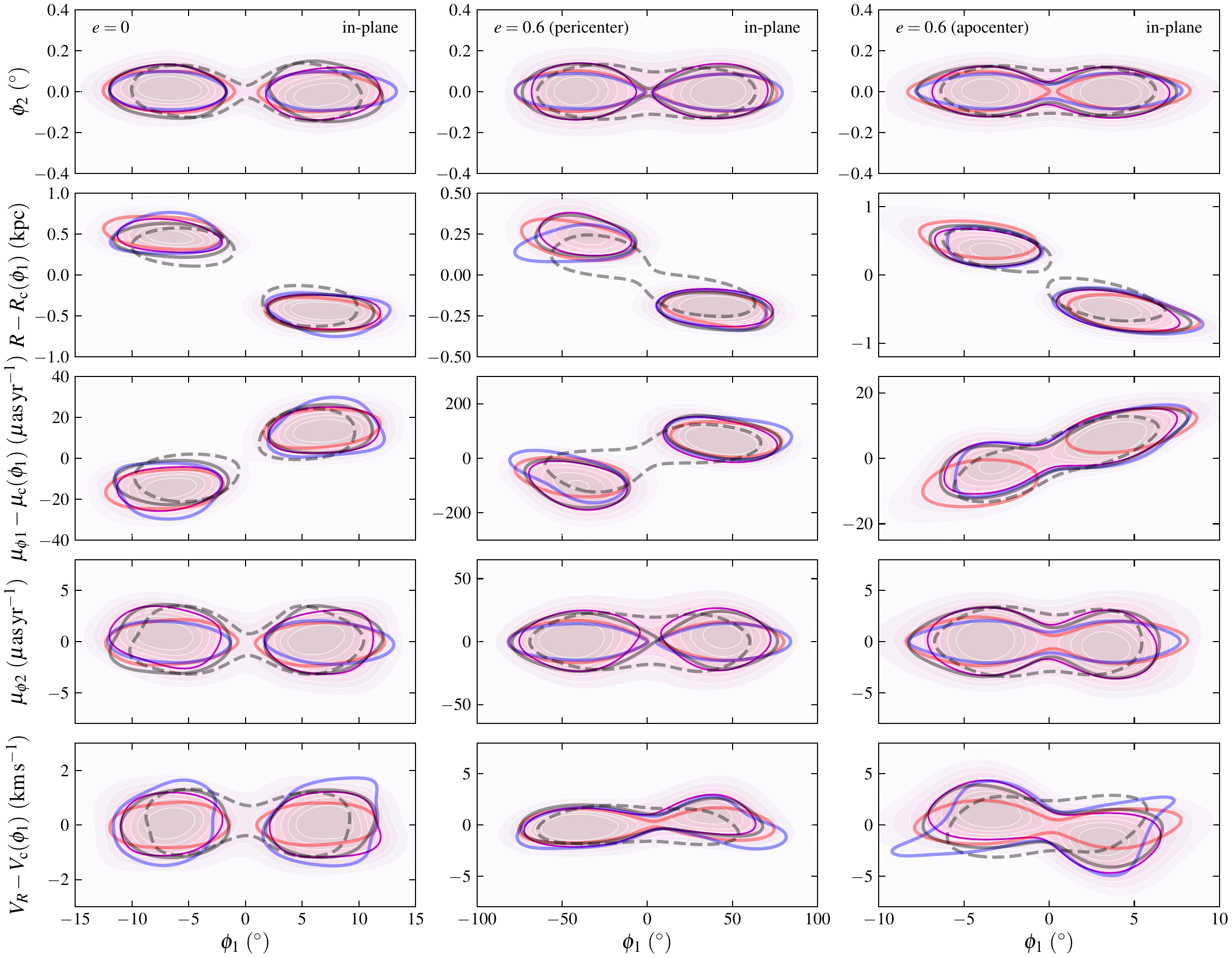}
    \caption{Stream morphology and velocities as in Fig.~\ref{fig:morphology_compare}, but evolved for another 500~Myr with the progenitor cluster removed.}
    \label{fig:morphology_compare_enlongation}
\end{figure*}

The defining feature of stellar streams is their elongated, stream-like distributions in the 2D celestial coordinate system and sometimes even the full 6D phase space. In this section, we test the ability of particle spray algorithms to reproduce these distributions as created by N-body simulations.

We investigate the distributions in the galactocentric frame, which is axis-aligned with the coordinate system in \S\ref{sec:distribution} such that the $X$--axis points at the progenitor GC. We apply spherical coordinates $(\phi_1,\phi_2,R)$, where $\phi_1$ is the angular coordinate along the cluster's velocity and $\phi_2$ is perpendicular to its velocity. $R$ is the radial distance towards the galactic center. Correspondingly, we describe the velocities using proper motions and the radial velocity $(\mu_{\phi 1},\mu_{\phi 2},V_{R})$. Such a coordinate system is similar to the great circle frame commonly used in observations to describe the morphology and velocities of stellar streams (except that the great circle frame is commonly constructed in the heliocentric frame), where $\phi_1$ is along the elongated direction of the stream.

In Fig.~\ref{fig:morphology_compare}, we plot the spatial and velocity distributions of escaped particles for in-plane orbits with $\ecc=0$ and $0.6$ from particle spray models and N-body simulations. From the top row to the bottom row, we plot $\phi_2$, $R$, $\mu_{\phi 1}$, $\mu_{\phi 2}$, and $V_R$ against the elongated direction $\phi_1$. Similarly to the previous section, we fix $(M_0,R_{\rm apo})=(10^5\Msun,40\ {\rm kpc})$ and only vary $\ecc$ and the orbital inclination. For the eccentric orbit $\ecc=0.6$, the stream morphology changes periodically at different orbital phases. We show two representative cases during the last pericenter/apocenter passage where the stream is the most/least elongated. For the circular orbit $\ecc=0$, we only show the distributions of these coordinates in the last snapshot as the stream morphology and kinematics do not change periodically with the orbital phase. Although $\phi_2$ and $\mu_{\phi 2}$ are always around zero in the galactocentric frame, $R$, $\mu_{\phi 1}$, and $V_R$ may vary across wide dynamical ranges. This makes it visually harder to distinguish the differences between models since such differences are typically much smaller than the dynamical ranges. Therefore, we subtract out the progenitor cluster's $R$, $\mu_{\phi 1}$, and $V_R$ along its orbit to only show the deviation of these coordinates from the unperturbed orbit. All distributions in Fig.~\ref{fig:morphology_compare} are represented as contours derived from Gaussian KDE.

The simulated streams have rather stable angular widths $\approx 0.1^\circ$, as measured by the standard deviation of $\phi_2$. For the circular ($\ecc=0$) case, the \citet{fardal_generation_2015} and \citet{roberts_stellar_2024} models slightly underestimate the widths of the streams. Specifically, these models produce a $15-30\%$ lower standard deviation of $\phi_2$ and $\mu_{\phi 2}$. Additionally, the \citet{fardal_generation_2015} model underestimates the radial velocity dispersion by about $30\%$, likely because it does not account for radial motion of the ejected particles. Conversely, the \citet{roberts_stellar_2024} model overestimates the radial velocity dispersion by approximately $40\%$. In contrast, our default model yields streams that are slightly wider, with a $20\%$ greater dispersion in $\phi_2$ and $\mu_{\phi 2}$ compared to N-body simulations. However, it accurately reproduces the radial velocity dispersion, with an error of less than $10\%$.

For the eccentric ($\ecc=0.6$) case, our default model provides the best match to simulations. In contrast, the \citet{fardal_generation_2015} and \citet{roberts_stellar_2024} models tend to generate streams that are too narrow in both $\phi_2$ and $\mu_{\phi 2}$. These models underestimate the standard deviation of these quantities by $30-50\%$, while our default model is accurate to within $3-7\%$ at both pericenter and apocenter. Additionally, the \citet{fardal_generation_2015} model produces a $40\%$ lower dispersion in $V_R$, whereas the \citet{roberts_stellar_2024} model overestimates it by $10-25\%$. Our default model, on the other hand, achieves an accuracy of $7\%$ for the radial velocity dispersion.

It is worth noting that the phase space distributions in our default model have standard deviations up to 30\% smaller than those from N-body simulations of eccentric orbits with $\ecc = 0.6$ (see Figs.~\ref{fig:spray_params_fixed_mean} and \ref{fig:spray_params_cluster_mass_orientation}). However, if we match the deviations exactly to the simulation values, the resulting streams become $10-20\%$ wider than the simulated ones. This discrepancy is likely because the phase space distributions of highly eccentric orbits cannot be perfectly captured by a multivariate Gaussian distribution. For instance, we notice that $\sigma_\theta$ actually decreases with the escape radius rather than a constant as in our model. Our model thus overestimates the stream width from these outer particles, which turn out to contribute more to the final width than the inner ones. While understanding and parameterizing the actual phase space distribution is beyond the scope of this work, we emphasize that our parameterization is already more detailed and sophisticated than most other particle spray models.

Additionally, we find that the deviations in phase space distributions contribute differently to the stream width and velocity dispersion. For example, increasing $\sigma_\theta$ by 20\% increases the stream width by around 10\%, while changes in $\sigma_\beta$ have an even smaller effect on the width. Other parameters do not significantly impact the stream width. Since our model deviations are $\lesssim 30\%$ smaller than those from the simulations, this corrects the overestimation of the stream's width and velocity dispersion for highly eccentric orbits.

Our ``no progenitor'' model produces standard deviation of the above coordinates with $5-20\%$ accuracy. While this is less accurate than our default model, it is still better than the \citet{fardal_generation_2015} and \citet{roberts_stellar_2024} models. However, the ``no progenitor'' model fails to accurately capture the distinct separation between the trailing and leading arms in $R$ and $\mu_{\phi 1}$, while all other models successfully reproduce this separation. Additionally, the this model tends to underestimate the length of the stream, with a $\sim10\%$ lower standard deviation of $\phi_1$. In contrast, the \citet{fardal_generation_2015} and \citet{roberts_stellar_2024} models overestimate this dispersion by $\sim10-30\%$. Our default model remains the most accurate for reproducing the standard deviation of $\phi_1$, with an error of only $3-7\%$.

Note that the length of the simulated stream increases over time. Specifically, the standard deviation of $\phi_1$ at 1~Gyr and 2~Gyr is roughly $1/3$ and $2/3$ of the final standard deviation at 3~Gyr. Our default model still performs the best at these earlier epochs, with relative errors similar to those observed at 3~Gyr.

We also compare the model outputs for polar orbits in Fig.~\ref{fig:morphology_compare_polar}. Compared to the in-plane orbits, the stream widths in $\phi_2$ for polar orbits are notably wider by a factor of 2. Our default model provides the best match to the spatial and velocity dispersion from N-body simulations, with errors around $\sim 10\%$. In contrast, the other models have errors up to $50\%$. This further demonstrates that a single set of spray parameters can effectively describe the morphology and velocities of streams across a wide range of orbital properties.

\subsection{Fully disrupted progenitor cluster}
\label{sec:fully_disrupted}

All above N-body simulations do not fully disrupt the progenitor GC. However, many observed stellar cluster streams, such as the GD-1 stream, are not associated with a surviving cluster. To assess the model's performance for fully disrupted clusters, we introduce an additional test case, where we manually remove the progenitor cluster at the end of the simulations, and then evolve the remaining particles for an additional 0.5~$\rm kpc\,km^{-1}\,s \approx 500$~Myr.

In Fig.~\ref{fig:morphology_compare_enlongation}, we compare the morphology and velocity distributions from simulations and particle spray models for in-plane orbits after the progenitor cluster has been removed. After this removal, the trailing and leading arms become more separated in the $\phi_1$ direction, as no new particles join the stream. The streams generated by the \citet{fardal_generation_2015} and \citet{roberts_stellar_2024} models are narrower than those from N-body simulations in the $\phi_2$ and $\mu_{\phi 2}$ directions. For the $\ecc=0.6$ case, the standard deviation of these quantities is underestimated by 30-50\%, and the error in the $V_R$ dispersion is similar. In contrast, our default model achieves an accuracy of about 10\% for these quantities. We also note that these results are consistent with those from the previous subsection, suggesting that the errors in the particle spray models evolve only weakly after the complete disruption of the progenitor clusters.

Since actions are conserved in the absence of the progenitor cluster (the gravity between escaped particles is negligible), the action space distributions remain nearly identical to those shown in Fig.~\ref{fig:action_comparison}. Therefore, we do not repeat the action space comparison here. Similarly, we do not show the results for polar orbits since they follow a similar pattern to the in-plane case.

\section{Summary}
\label{sec:summary}

In this work, we develop a new particle spray model for numerically modeling GC streams by directly drawing the initial positions and velocities of escaped stream particles from a multivariate Gaussian distribution. The parameters of this distribution are calibrated to match N-body simulations. We first run a suite of N-body simulations of GCs orbiting a galactic potential, varying parameters such as GC mass, orbital radius, eccentricity, and inclination (see Table~\ref{tab:nbody}). The progenitor clusters are initialized using a King model with $W=8$, which is derived from the observed tidal radius--core radius relation for MW GCs (see Fig.~\ref{fig:r_tid_r_core}).

We define the escape time of a simulation particle as the moment when its velocity reaches the escape velocity from the GC and subsequently remains above it. By recording the particles' positions and velocities in a spherical coordinate system centered on the GC, we observe that the five independent phase space coordinates $(r, \phi, \theta, \alpha, \beta)$ of newly escaped particles follow a multivariate Gaussian distribution (see Fig.~\ref{fig:spray_params_fixed_mean_1d}), with an anti-correlation only between the $r$--$\alpha$ pair (see Fig.~\ref{fig:spray_params_fixed_mean}). This anti-correlation is also illustrated in Fig.~\ref{fig:escape_velocity}.

We find that $R_{\rm apo}$ and $M_0$ have little effect on the phase space distribution. Although the mean values of the distribution depend weakly on the orbital eccentricities $\ecc$, the variation of the phase space distribution increase with $\ecc$ (see Fig.~\ref{fig:spray_params_fixed_mean}). The orbital inclination also affects the variations in the distribution. Additionally, polar orbits have greater $\overline{r}$ compared to in-plane orbits (see Fig.~\ref{fig:spray_params_cluster_mass_orientation}). However, these are only minor changes that do not significantly affect the general shapes of the distributions. 

Since the dependence on orbital properties is weak, we approximate the distribution function of newly-escaped particles using a universal set of parameters for the multivariate Gaussian distribution, as defined in Eqs.~(\ref{eq:gaussian}) to (\ref{eq:parameter}). Based on this approximation, we develop a new particle spray algorithm that draws the initial positions and velocities of escaped particles from this multivariate Gaussian distribution and evolves these particles solely under the joint potential of the host galaxy and the progenitor cluster.

Next, we perform a series of tests to evaluate the performance of the new algorithm. Compared to existing algorithms such as \citet{fardal_generation_2015} and \citet{roberts_stellar_2024}, the new algorithm provides a significantly better match to the final action space distribution from N-body simulations (Fig.~\ref{fig:action_comparison}). Furthermore, the new model accurately reproduces the stream morphology and kinematics from N-body simulations, achieving a stream width in the phase space accurate to about 10\%, while other methods tend to underestimate it by up to 50\% (Figs.~\ref{fig:morphology_compare} and \ref{fig:morphology_compare_polar}). Also, our new algorithm outperforms other methods for fully disrupted GCs, where we evolve the stream particles for an additional 500~Myr after removing the progenitor cluster (Fig.~\ref{fig:morphology_compare_enlongation}).

We also conducted tests with various numerical parameters, including the $W$ parameter for the King model, time step, softening length, and particle mass. Although the mass loss rate can vary significantly with different settings (Fig.~\ref{fig:massloss_action_test}), the phase space distributions are much less sensitive to these parameters (Fig.~\ref{fig:spray_params_test}). The only exception is the case with $W=2$, where the GC becomes so diffuse that its half-mass radius approaches the tidal radius. Such a diffuse profile is more typical of dwarf galaxies, highlighting the need for a particle spray model specifically optimized for GC streams.

For public use, we have implemented the new particle spray model in galactic dynamics codes \texttt{agama}, \texttt{gala}, \texttt{galax}, and \texttt{galpy}. Online notebooks detailing how to use these implementations are available  via\dataset[DOI: 10.5281/zenodo.13923250]{https://doi.org/10.5281/zenodo.13923250} and at \url{https://github.com/ybillchen/particle_spray}.

\section*{Acknowledgements}
We are grateful to Eric Bell, Ana Bonaca, Jiang Chang, Raphael Errani, Sergey Koposov, Guoliang Li, Hui Li, Jacob Nibauer, Carles Garcia Palau, Adrian Price-Whelan, Behzad Tahmasebzadeh, Eugene Vasiliev, Wenting Wang, and Newlin Weatherford for insightful discussions. In particular, we extend special thanks to Eugene Vasiliev, Adrian Price-Whelan, Nathaniel Starkman, and Jo Bovy for their thorough review of the code implementing our new algorithm in \texttt{agama}, \texttt{gala}, \texttt{galax}, and \texttt{galpy}.
We also thank the reviewer for their suggestions and comments, which have improved the quality of this work. OG and YC were supported in part by the U.S. National Science Foundation through grant AST-1909063 and by National Aeronautics and Space Administration (NASA) through contract NAS5-26555 for Space Telescope Science Institute program HST-AR-16614. MV and NA gratefully acknowledge financial support from NASA-ATP awards  80NSSC20K0509 and 80NSSC24K0938. 
This research benefited from the Dwarf Galaxies, Star Clusters, and Streams Workshop hosted by the Kavli Institute for Cosmological Physics.

\software{
\texttt{numpy} \citep{harris_array_2020}, 
\texttt{matplotlib} \citep{hunter_matplotlib_2007}, 
\texttt{scipy} \citep{virtanen_scipy_2020}, 
\texttt{agama} \citep{vasiliev_agama_2019}, 
\texttt{falcON} \citep{dehnen_very_2000,dehnen_hierarchical_2002}, 
\texttt{PeTar} \citep{wang_petar_2020}, 
\texttt{galpy} \citep{bovy_galpy_2015}
}

\newpage
\appendix
\vspace{-6mm}

\section{Effects of the $W$ parameter}
\label{sec:W_test}

After fixing the outer radius to the tidal radius of star clusters, there is only one free parameter for the King model, the dimensionless central potential $W$. Although the average $r_{\rm tid}$--$r_{\rm core}$ ratio suggests $W \approx 8$ for MW GCs (Fig.~\ref{fig:r_tid_r_core}), this parameter actually ranges between 4 and 12. To assess how different values of $W$ affect key results such as the initial positions, initial velocities, and the action space distribution of escaped particles, we conduct a series of test runs with $W$ ranging from 2 to 14. For this test, we use a representative simulation setting: $(M_0, R_{\rm apo}, \ecc) = (10^5 \Msun, 40 \ {\rm kpc}, 0.6)$ with an in-plane orbit.

\begin{figure*}
    \centering
    \includegraphics[width=\linewidth]{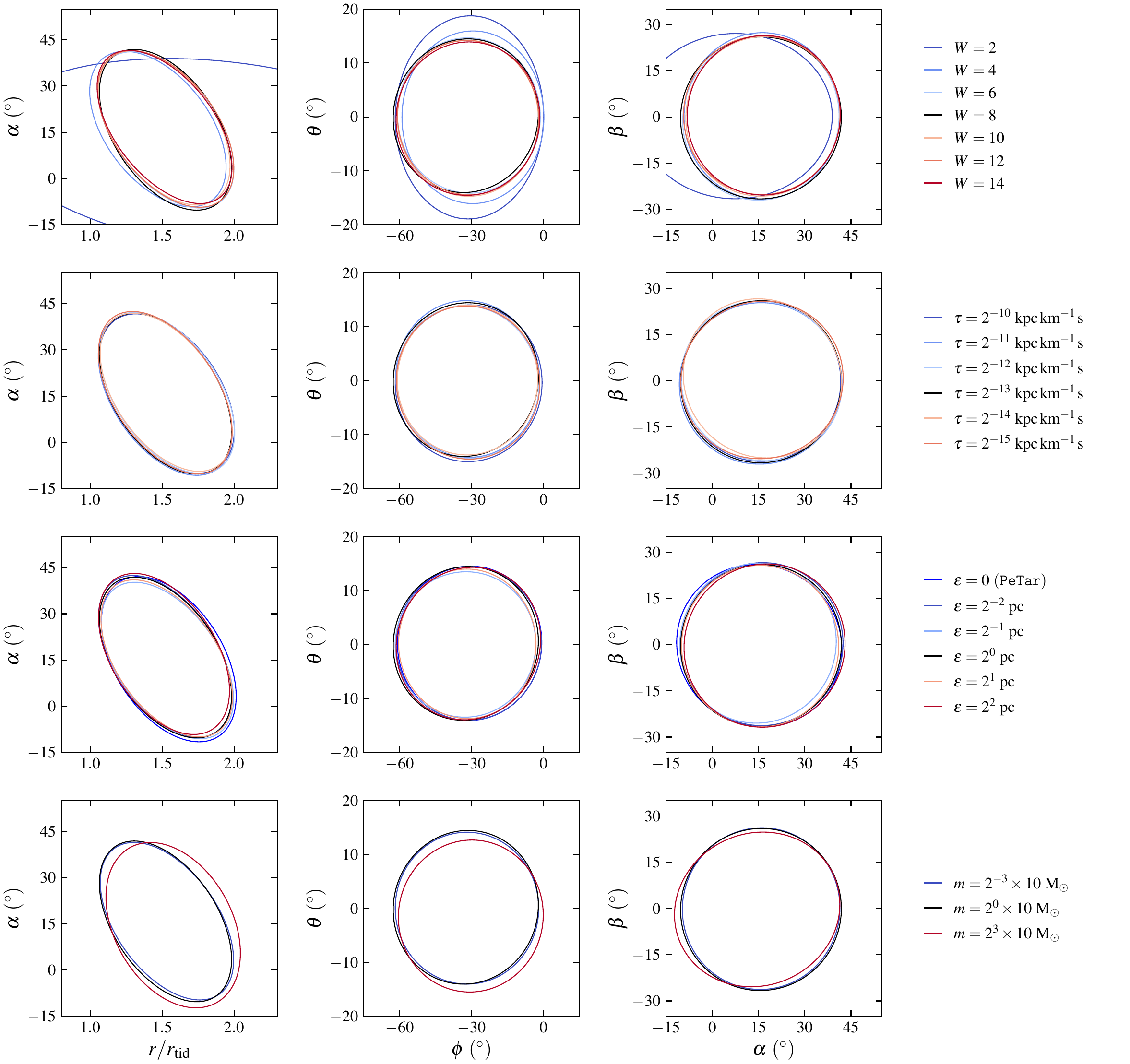}
    \caption{Initial position and velocity distributions for escaped particles as in Fig.~\ref{fig:spray_params_fixed_mean}, but varying the $W$ parameter for the King model (\textit{first row}),  time step $\tau$ (\textit{second row}), softening length $\varepsilon$ (\textit{third row}), and particle mass $m$ (\textit{fourth row}). The physical parameters are $(M_0, R_{\rm apo}, \ecc)=(10^5\Msun,40\ {\rm kpc}, 0.6)$, with the in-plane orbital inclination.}
    \label{fig:spray_params_test}
\end{figure*}

In the \textit{first row} of Fig.~\ref{fig:spray_params_test}, we show the initial phase space distributions of escaped particles. A wide range of $W = 4$ to $14$ does not significantly alter these distributions, while the $W = 2$ case has a much larger spread in $r$ and $\theta$ and a shift in $\alpha$ towards zero. This is likely due to the cluster being so diffuse that its half-mass radius is comparable to the tidal radius. Additionally, as the cluster approaches pericenter, its tidal radius can shrink below the half-mass radius. Particles outside the tidal radius quickly escape with significant dispersion in both their position and velocity distributions. Consequently, the outskirts of this cluster are more sensitive to the tidal field and escape in a more stochastic fashion. This scenario is more similar to the streams that emerge from dwarf galaxies, highlighting the need for a particle spray model specifically optimized for GC streams.

\begin{figure*}
    \centering
    \includegraphics[width=0.75\linewidth]{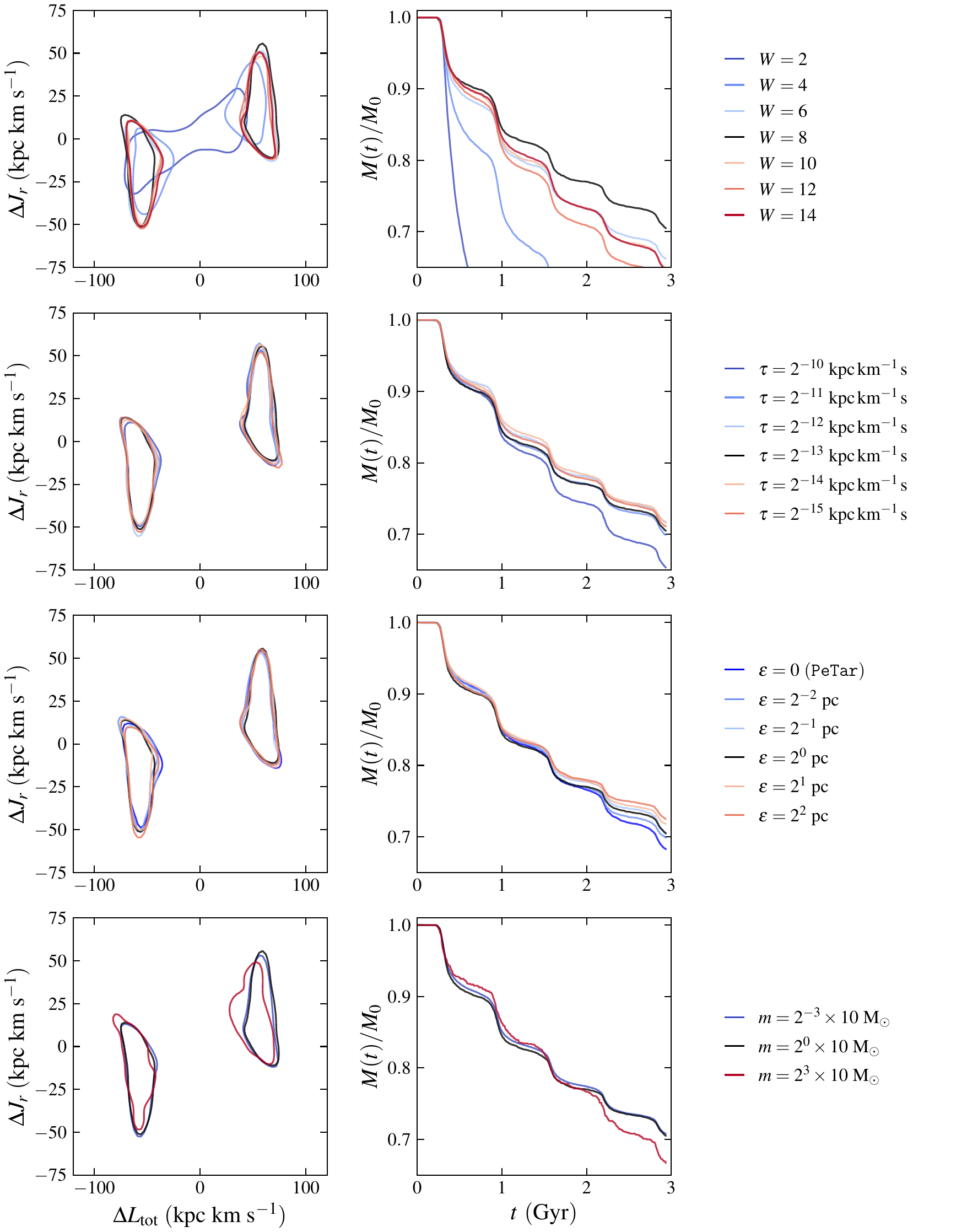}
    \caption{Action space distributions of stream particles (\textit{left}) and mass loss histories of the progenitor cluster (\textit{right}), varying the $W$ parameter for the King model (\textit{first row}),  time step $\tau$ (\textit{second row}), softening length $\varepsilon$ (\textit{third row}), and particle mass $m$ (\textit{fourth row}). The physical parameters are $(M_0, R_{\rm apo}, \ecc)=(10^5\Msun,40\ {\rm kpc}, 0.6)$, with the in-plane orbital inclination. Similarly to Fig.~\ref{fig:action_comparison}, the contours in the \textit{left column} enclose 50\% of total particles.}
    \label{fig:massloss_action_test}
\end{figure*}

However, different values of $W$ lead to distinct mass loss rates for the progenitor cluster, as shown in the \textit{right panel} of the \textit{first row} of Fig.~\ref{fig:massloss_action_test}. The $W = 8$ case has the lowest mass loss rate, likely because it has the smallest half-mass radius-to-tidal radius ratio among all $W$ values considered, even though its core radius-to-tidal radius ratio is not the smallest. This suggests that the $W = 8$ case has the most concentrated mass profile and is less affected by tidal disruption, consistent with previous studies \citep[e.g.,][]{johnstone_evaporation_1993, gnedin_destruction_1997}. In contrast, the $W = 2$ case represents the least concentrated profile, with a mass loss rate so high that it is completely disrupted within the first 2~Gyr. Consequently, the action space distribution of these particles at the last snapshot (\textit{left panel} of the \textit{first row} of Fig.~\ref{fig:massloss_action_test}) resembles the ``no progenitor'' case shown in Fig.~\ref{fig:action_comparison}. Despite $W = 2$, the action space distributions for all other $W$ values remain similar.

In conclusion, although different $W$ affect the mass loss rate of the progenitor cluster, a wide range of $W = 4$ to $14$ does not significantly alter the initial phase space distributions for escaped particles. This robustness of the distributions enables us to parameterize the particle spray model with a universal set of parameters.

\section{Convergence tests}
\label{sec:convergence}

In this section, we perform a set of convergence tests to assess how simulation parameters impact the results. We test three numerical parameters: time step $\tau$, softening length $\varepsilon$, and particle mass $m$. These parameters should not affect the collisionless physical processes as long as they are within appropriate ranges.

\subsection{Time step}

Ideally, the time step should be smaller than the softening length crossing time, $\tau \lesssim \varepsilon / \sigma$, where $\sigma$ is the velocity dispersion within the star cluster. In practice, a shorter $\tau$ results in longer runtime, and a time step that is too short does not improve accuracy further. Thus, it is crucial to balance accuracy and efficiency by choosing an appropriate $\tau$.

To examine the simulation convergence with respect to the time step, we vary $\tau$ for the same test case as described in Appendix~\ref{sec:W_test}: $(M_0, R_{\rm apo}, \ecc) = (10^5 \Msun, 40\ {\rm kpc}, 0.6)$ with in-plane orbits. In addition to the fiducial numerical parameters $(\tau, \varepsilon, m) = (2^{-13}\ {\rm kpc\,km^{-1}\,s}, 1\,{\rm pc}, 10\Msun)$, which are the same as those used in this work, we perform five test runs with $\tau = 2^{-15} - 2^{-10}\ {\rm kpc\,km^{-1}\,s} \approx 0.03 - 1$~Myr. For reference, the softening length crossing time is approximately $0.3$~Myr with $\sigma \approx 3 \kms$.

We investigate the phase space distributions of newly-escaped particles in the \textit{second row} of Fig.~\ref{fig:spray_params_test}. Varying $\tau$ does not result in any noticeable change in these distributions. We also examine the action space distribution of all escaped particles at the final snapshot, shown in the \textit{left panel} of the \textit{second row} of Fig.~\ref{fig:massloss_action_test}. Again, all test runs yield similar distributions.

The mass loss history, on the other hand, depends on $\tau$. As shown in the \textit{top right panel} of the \textit{second row} of Fig.~\ref{fig:massloss_action_test}, the mass loss rate does not converge and increases with $\tau$ until $\tau \lesssim 2^{-11}\ {\rm kpc\,km^{-1}\,s} \approx 0.5$~Myr. It stabilizes when $\tau$ is smaller than this threshold value, with the final mass of the star cluster settling at $(0.70-0.72)\,M_0$. The variation is likely due to intrinsic randomness, as re-running the simulation with the same setup but a different random seed can also lead to similar variations. To ensure numerical convergence for N-body simulations, we choose a safe value of $2^{-13}\ {\rm kpc\,km^{-1}\,s} \approx 0.1$~Myr throughout this paper.

\subsection{Softening length}
\label{sec:softening}

We normally expect the softening length to be greater than the average separation between particles and smaller than the size of the structure we wish to resolve (i.e., the core radius of the cluster) in the collisionless regime. For the default settings, this condition roughly translates to $0.2\ {\rm pc} \lesssim \varepsilon \lesssim 3$~pc. We examine the simulation dependence on the softening length by performing four additional test runs with $\varepsilon = 2^{-2} - 2^{2}$~pc. To match the condition $\tau \lesssim \varepsilon / \sigma$, we also scale $\tau$ proportionally to $\varepsilon$.

Additionally, we compare the results with zero softening length ($\varepsilon=0$), which accounts for close collisional encounters between particles. For this comparison, we use the collisional N-body code \texttt{PeTar} (\citealp{wang_petar_2020}, built upon \citealp{iwasawa_implementation_2016,namekata_fortran_2018,wang_slow-down_2020}), which incorporates the same \texttt{MWPotential2014} galactic potential as in our simulations.

In the \textit{third row} of Fig.~\ref{fig:spray_params_test}, we show that the phase space distributions of the newly-escaped particles are not significantly affected by varying softening lengths. Similarly, the final action space distribution is only weakly dependent on $\varepsilon$, as illustrated in the \textit{left panel} of the \textit{third row} of Fig.~\ref{fig:massloss_action_test}.

We also examine the mass loss histories for different softening lengths, shown in the \textit{right panel} of the \textit{third row} of Fig.~\ref{fig:massloss_action_test}. The final remaining mass increases from $0.68\,M_0$ to $0.73\,M_0$ as the softening length $\varepsilon$ changes from 0 to 4~pc. Notably, there is no significant incontinuity between the collisional case ($\varepsilon=0$) and cases with $\varepsilon > 0$. 

For this work, we select an intermediate value of $\varepsilon=1$~pc. However, the specific choice of $\varepsilon$ is less critical, as it has a minimal impact on the initial phase space distributions, the final action space distribution, and the mass loss history of GC streams.

\subsection{Particle mass}

In the collisionless regime, the mass of individual particles should be much smaller than the total mass of the resolved structure (i.e., the mass of the cluster). Although there is no strict lower limit for particle mass as long as computational resources permit, setting the particle mass too low may create artificial structures even smaller than the smallest elements involved in gravitational interaction (e.g., individual stars). Resolving such unphysical structures is unnecessary. Therefore, the ideal particle mass should satisfy $\Msun \lesssim m \ll M_0$.

In addition to the default particle mass of $10\Msun$, we test two additional masses: $1.25\Msun$ and $80\Msun$, which are $2^{-3}$ and $2^{3}$ times the default mass, respectively. Consequently, the average separation between particles changes to $2^{-1}$ and $2$ times the default separation. To maintain $\varepsilon$ as greater than the particle separation by the same ratio, we adjust the softening length by $2^{-1}$ and $2$, respectively. Similarly, the time step is scaled proportionally with $\varepsilon$ to ensure the condition $\tau \lesssim \varepsilon / \sigma$.

Similar to the previous tests, the phase space and action space distributions are only weakly dependent on particle mass, as shown in the \textit{fourth row} of Fig.~\ref{fig:spray_params_test} and the \textit{left panel} of the \textit{fourth row} of Fig.~\ref{fig:massloss_action_test}. However, the case with $m = 2 \times 10^3 \Msun$ exhibits a higher mass loss rate, as indicated in the \textit{right panel} of the \textit{fourth row} of Fig.~\ref{fig:massloss_action_test}. This increased mass loss is likely due to numerical heating when the number of particles is too low. Conversely, reducing the particle mass further does not significantly affect the mass loss rate compared to the default case, suggesting that the simulation results have already converged with the default particle mass of $10 \Msun$.

Finally, we conclude that our numerical setup $(\tau, \varepsilon, m) = (2^{-13}\ {\rm kpc\,km^{-1}\,s}, 1\ {\rm pc}, 10\Msun)$ is appropriate to minimize numerical effects that lead to artificial errors. This setup balances accuracy and computational efficiency by preventing unnecessarily small $\tau$ and $m$. Furthermore, we find that the initial phase space distributions of escaped particles are more robust to changes in numerical parameters than the mass loss rate. Therefore, while our simulations may not perfectly capture the mass loss histories of star clusters, our parameterization of the phase space distributions remains reliable across different numerical settings.

\bibliography{GC-model-references}
\bibliographystyle{aasjournal}

\end{document}